\documentclass[12pt]{article}
\begin{document}
\begin{titlepage}
\begin{flushright}
TUM-HEP-617/06 \\
KEK-TH-1063
\end{flushright}
\begin{center}
{\Large\bf 
New contribution to dimension five operators 
 on proton decay in anomaly mediation scenario 
}  
\end{center}
\vspace{1cm}
\begin{center}
Naoyuki {Haba}$^{(a),}$ 
\footnote{E-mail: haba@ph.tum.de}
and 
Nobuchika {Okada}$^{(b),}$
\footnote{E-mail: okadan@post.kek.jp} 
\end{center}
\vspace{0.2cm}
\begin{center}
${}^{(a)}${\it Physik-Department, 
 Technishe Universit$\ddot{a}$t M$\ddot{u}$nchen, James-Franck Strasse,
 D-85748 Garching, Germany}
\\[0.2cm]
${}^{(b)}$ {\it Theory Division, KEK, Tsukuba 305-0801, 
Japan}
\\[0.2cm]
${}^{(c)}$ {\it Department of Particle and Nuclear Physics, 
The Graduate University \\ 
for Advanced Studies (Sokendai), 
Oho 1-1, Tsukuba 305-0801, Japan.}

\end{center}
\vspace{1cm}
\begin{abstract}

In supergravity, 
 effective superpotential relevant to dimension five operators 
 on proton decay processes also leads to 
 supersymmetry breaking terms among sfermions, 
 dimension four operators. 
These dimension four operators induce the dimension five operators 
 through 1-loop diagrams dressed by gauginos. 
We find that, 
 in a class of models with the anomaly mediation, 
 the 1-loop contributions can be comparable 
 to those at the tree level. 
Therefore, such operators have a great impact on proton decay rate. 
Depending on a universal phase of gaugino masses and soft mass spectrum, 
 the proton decay rate can be enhanced or suppressed. 

\end{abstract}
\end{titlepage}
\newpage

Supersymmetry (SUSY) extension is one of the most promising ways 
 to provide a solution to the gauge hierarchy problem 
 in the Standard Model \cite{review}. 
The minimal version of SUSY extension of the Standard Model
 is so-called the Minimal Supersymmetric Standard Model (MSSM). 
Interestingly, the experimental data support 
 the unification of the three gauge couplings in the MSSM 
 at the scale $M_{GUT} \sim 2 \times 10^{16} \mbox{GeV}$ 
 \cite{unification}. 
At high energies, our world may be described 
 by a SUSY grand unified theory (GUT) 
 with a simple gauge group such as $SU(5)$, $SO(10)$, $E_6$ etc. 
 into which all the gauge groups in the standard model 
 are embedded and unified. 

The most characteristic prediction of the (SUSY) GUTs 
 is the proton decay. 
Normally in SUSY GUTs the proton decay process 
 through the dimension five operators \cite{dim5} 
 involving MSSM matters and the color triplet Higgsino 
 turns out to be the dominant decay modes, 
 since the process is suppressed by only a power of 
 the color triplet Higgsino mass scale. 
Experimental lower bound on the proton decay modes 
  $p \to K^{+} \overline{\nu}$ through the dimension five operators 
  is given by SuperKamiokande \cite{sk}, 
\begin{eqnarray}
\tau (p \rightarrow K^{+} \overline{\nu}) \geq 
 2.2 \times 10^{33} \,\, [{\mathrm{years}}]  \nonumber .  
\end{eqnarray}
This is one of the most stringent constraints on SUSY GUT models. 
In fact, it is widely believed that 
 the minimal SUSY $SU(5)$ model is excluded 
 from the experimental bound 
 together with the requirement of the success  
 of the three gauge coupling unification \cite{gn} \cite{mp}. 

However, for the proton decay processes 
 through the dimension five operators, 
 knowledge of the flavor structure is essential 
 in order to give definite predictions for proton decay rate.
Some models in which flavor structures are extended 
 have been found to be consistent with the experiments 
 \cite{bps} \cite{ew}. 
Also, if a GUT model has an extended structure for the Higgs sector, 
 the proton decay rate depends on free parameters in 
 the extended Higgs sector and, together with flavor structures, 
 some model can avoid the stringent bound on the proton life time 
 \cite{fikmo}. 
Although rapid proton decay would be avoidable in such ways, 
 we need very complicated studies. 
It must be interesting if there exists new contribution 
 to proton decay processes 
 which can reduce proton decay rate 
 without involving such complicated structures. 

In this letter, we show that there exists such 
 a new contribution to proton decay processes 
 through dimension five operators, 
 when SUSY breaking mediation obeys 
 the anomaly mediated SUSY breaking (AMSB) scenario  
 \cite{AMSB1} \cite{AMSB2}. 
In a class of models with the AMSB, 
 SUSY breaking order parameter for superpartners 
 in the MSSM is the mass of gravitino, 
 more correctly, F-term of the compensating multiplet 
 in superconformal framework of supergravity 
 \cite{superconformal1} \cite{superconformal2} 
\footnote{ 
 This fact is crucial in some SUSY breaking models. 
 Although the F-term of the compensating multiplet is 
 of the same order of the gravitino mass in a simple 
 SUSY breaking scenario, 
 it is not a general consequence in supergravity. 
 Gravitino can, in general, be super heavy, leaving the AMSB 
 contribution to the soft SUSY mass spectrum in the MSSM 
 around the electroweak scale \cite{almost}. 
}. 
Then, the scale of the gravitino mass is roughly 
 two orders of magnitude larger than that of 
 soft SUSY breaking masses induced by the AMSB. 

In supergravity, any dimensionful parameter in a model
 is always accompanied by the compensating multiplet, 
 so that, once SUSY is broken, a SUSY breaking term 
 proportional to the gravitino mass appears. 
A simple example is the $B$-term for 
 the Higgs doublets in the MSSM. 
If we introduce $\mu$-term in Higgs superpotential, 
 we obtain $B \sim \mu m_{3/2}$
\footnote{ 
In AMSB scenario, this B-term is too big, 
 because of the large gravitino mass. 
Thus, we need a fine-tuning to obtain 
 the correct scale of the electroweak symmetry breaking. 
For an idea to reduce the B-term, 
 see, for example, Ref.~\cite{PR}.  
} 
after SUSY breaking, 
 where $m_{3/2}$ is the gravitino mass. 
The similar thing happens even to higher dimensional operators. 
This fact can be easily understood in the following way. 
Imagine that a higher dimensional operator is the result 
 after integrating out some heavy fields. 
Thus, soft SUSY breaking term for the higher dimensional 
 operator is originating from $B$-terms 
 for the heavy fields integrated out. 
We examine such soft SUSY breaking terms originating 
 from superpotentials which leads to the dimension five operators 
 relevant to the proton decay. 
We will show that such terms have a great impact 
 on the resultant proton decay rate. 

For simplicity, we work on the minimal SUSY $SU(5)$ model. 
Application to other GUT models is straightforward. 
In the minimal SUSY $SU(5)$ model, effective superpotential 
 obtained after integrating out color triplet Higgs superfields 
 is given by \cite{hisano} 
\begin{eqnarray} 
W_{eff} =  \frac{1}{2 M_{H_C}} Y_{ikl} 
 \left( Q_i Q_i \right) \left( Q_k L_l \right)  
+ \frac{1}{M_{H_C}} Y_{ijkl} 
 \left( u^c_i e^c_j \right) \left( u^c_k d^c_l \right)  , 
 \label{eff} 
\end{eqnarray} 
where $M_{H_C}$ is a supersymmetric mass of the color 
 triplet Higgs superfields, which should be around the GUT scale, 
 and $Y_{ikl}$ and $Y_{ijkl}$ are some products of Yukawa couplings 
 (we do not need their explicit forms in the following discussion). 
Here, the contraction of the indices are understood as 
\begin{eqnarray} 
\left( Q_i Q_i \right) \left( Q_k L_l \right) 
 &=& \epsilon_{\alpha \beta \gamma} 
   \left( 
   u^\alpha_i d_i^{\prime \beta} 
   - d_i^{\prime \alpha} u_i^{\beta} \right) 
   \left( u^\gamma_k e_l - d_k^{\prime \gamma} \nu_l 
 \right), \\ \nonumber 
\left( u^c_i e^c_j \right) \left( u^c_k d^c_l \right)  
 &=& \epsilon^{\alpha \beta \gamma}  
 u^c_{i \alpha} e^c_j u^c_{k \beta} d^c_{l \gamma} ,   
\end{eqnarray} 
where $\alpha$, $\beta$, $\gamma$ are color indices  
 and $d^{\prime}_i$ is defined as $ d^{\prime}_i = V_{ij} d_j$ 
 with $V_{ij}$ being the Kobayashi-Maskawa matrix. 
Extracting two fermions and two sfermions 
 from the effective Lagrangian, 
 we obtain dimension five operators relevant to proton decay. 

Now let us see SUSY breaking effects on the effective 
 superpotential. 
It is useful to work out in the superconformal framework 
 of supergravity \cite{superconformal2}. 
In this framework, SUSY breaking is encoded in F-component  
 of the compensating multiplet $\phi = 1+ \theta^2 F_\phi$.  
Normal SUSY breaking models lead to $F_\phi \sim m_{3/2}$ 
 due to the condition for vanishing cosmological constant 
 (vacuum energy). 
In the original basis in the superconformal framework, 
 the basic Lagrangian for matter and Higgs chiral superfields   
 is symbolically written as 
\begin{eqnarray} 
 {\cal L}= 
 \int d^4 \theta \phi^\dagger \phi {\cal K} 
 + \left( \int d^2 \theta \phi^3 W  + h.c. \right),   
\end{eqnarray} 
where ${\cal K}$ and $W$ are Kahler potential and superpotential, 
 respectively. 
Assuming canonical Kahler potentials and rescaling all chiral
 superfields such as $X_i \rightarrow X_i/\phi$, 
 we can eliminate the compensating multiplet 
 from the above Kahler potential. 
The effective superpotential we are considering is the higher 
 dimensional one and  there exists the mass parameter $M_{H_C}$. 
Therefore, the compensating multiplet is left un-eliminated 
 in the superpotential. 
Resultant effective superpotential is found to be 
\begin{eqnarray} 
 \int d^2 \theta \frac{1}{2 \phi M_{H_C}} 
 Y_{ikl}  \left( Q_i Q_i \right) \left( Q_k L_l \right)  
+ \frac{1}{\phi M_{H_C} } Y_{ijkl} 
 \left( u^c_i e^c_j \right) \left( u^c_k d^c_l \right)  . 
\end{eqnarray} 
Substituting $ \phi=1 +\theta^2 F_\phi $, 
 this superpotential leads to soft SUSY breaking terms 
 among sfermions, 
\begin{eqnarray} 
  {\cal L}_{soft} \supset 
 -\frac{F_\phi}{2 M_{H_C}}  
  Y_{ikl}  \left( \tilde{Q}_i \tilde{Q}_i \right) 
  \left( \tilde{Q}_k \tilde{L}_l \right)  
 - \frac{F_\phi}{M_{H_C} } Y_{ijkl} 
  \left( \tilde{u^c}_i {\tilde e^c}_j \right) 
  \left( \tilde{u^c}_k \tilde{d^c}_l \right) 
 + h.c. . 
 \label{dim4}
\end{eqnarray} 
These are dimension four operators. 

In the above calculations, we used the effective superpotential 
 obtained by integrating out the color triplet Higgs superfields 
 under SUSY vacuum conditions 
 and plugged it into the superconformal framework. 
Of course, we can do the same calculations by using 
 the original superpotential before integrating the color 
 triplets out. 
In the original Lagrangian, we obtain soft SUSY breaking term, 
 $B$-term, for the color triplet Higgs bosons such as $F_\phi M_{H_C}$. 
After integrating out the color triplet Higgs bosons and Higgsinos, 
 we come to the same result as the above. 
However, correctly speaking, 
 there exist higher order terms with 
 respect to the expansion parameter, $(F_\phi/M_{H_C})$,  
 in actual calculations. 
The result in Eq.~(\ref{dim4}) gives the correct leading term. 
Through the calculations based on the original Lagrangian, 
 we can see that the dimension four operators originate 
 from the $B$-term of the color triplet Higgs bosons. 

Note that the soft SUSY breaking terms with dimension four 
 lead to dimension five operators through 1-loop diagram 
 dressed by gauginos.\footnote{ 
In the context of usual gravity mediated SUSY breaking 
in supergravity, 1-loop diagrams similar to 
those we consider in this paper 
were examined many years ago \cite{sakai}. 
We would like to thank Pavel Fileviez Perez 
for leading our attention to the paper. 
In the usual case, contribution from the 1-loop diagrams 
is subdominant, because the gravitino mass is 
around the weak scale. 
}
These dimension five operators have the same coefficients,    
 $Y_{ijk}$ and $Y_{ijkl}$, 
 as the ones obtained from Eq.~(\ref{eff}), 
 because the dimension four operators originate 
 the same effective superpotential. 
Here, as an example, 
 let us evaluate a dimension five operator 
 $(\psi_{Q_i} \psi_{Q_i}) (\tilde{Q}_k \tilde{L}_l)$ 
 induced by the dimension four operator in Eq.~(\ref{dim4}) 
 through 1-loop diagram dressed by gauginos. 
Calculations for other 1-loop diagrams are straightforward. 
For 1-loop diagram dressed by gluino, we obtain 
\begin{eqnarray}
\frac{4}{3} 
 \left( {\alpha_3 \over 4 \pi}  \right)
 \left( {F_{\phi}\over 2 M_{H_C}} \right)
 M_3^* 
 &\times&  f(M_3, m_{\tilde{u}_i}, m_{\tilde{d}_i}), \\
f(M_3, m_{\tilde{u}_i}, m_{\tilde{d}_i}) &=& 
\int_0^\infty 
 \frac{ k_E^2 d k_E^2 }
{\left(k_E^2 +|M_3|^2 \right)
 \left(k_E^2 + m_{\tilde{u}_i}^2  \right)
 \left(k_E^2 + m_{\tilde{d}_i}^2  \right)} \nonumber \\ 
&=& 
{1 \over (m_{\tilde{u}_i}^2- m_{\tilde{d}_i}^2)}
\left(
{m_{\tilde{u}_i}^2 \over m_{\tilde{u}_i}^2-|M_3|^2} 
\ln{m_{\tilde{u}_i}^2 \over |M_3|^2}-
{m_{\tilde{d}_i}^2 \over m_{\tilde{d}_i}^2-|M_3|^2} 
\ln{m_{\tilde{d}_i}^2 \over |M_3|^2}
\right), \nonumber \\ 
&\sim&
{1 \over (|M_3|^2- \tilde{m}^2)^2}
\left(
{\tilde{m}^2 -|M_3|^2-|M_3|^2  \ln{\tilde{m}^2 \over |M_3|^2}}
\right), 
\end{eqnarray}
where $\alpha_3$ is the QCD gauge coupling 
 (from fermion-sfermion-gaugino vertex), 
 $M_3$ is generally complex gluino mass, 
 and $m_i$ denotes a mass squared of squark running in a loop. 
We took the degenerate mass spectrum for 
 $m_{\tilde{u}_i} \sim m_{\tilde{d}_i} \equiv \tilde{m}$ 
 in the last line. 
The loop function $f$ is positive and 
 its approximation form in some limits is given by 
\begin{eqnarray}
 f(M_3, \tilde{m}) &\sim& 
  {1 \over \tilde{m}^2} \;\;\;\;\;\; (\mbox{for } |M_3| \ll \tilde{m}), \\
&\sim& {1\over |M_3|^2 } \left( 
 \ln \frac{|M_3|^2}{\tilde{m}^2} - 1 \right) 
   \;\;\; (\mbox{for } |M_3| \gg \tilde{m}).   
\end{eqnarray}
We obtain similar results for 1-loop corrections 
 dressed by wino and bino. 
In the limit of absence of mixing with Higgsinos, 
 1-loop corrections dressed by wino and bino are
 found to be 
\begin{eqnarray}
 \frac{3}{2} 
 \left( {\alpha_2 \over 4 \pi} \right) 
 \left( {F_{\phi}\over 2 M_{H_C}} \right)
 M_2^* &\times&  f(M_2, m_{\tilde{u}_i}, m_{\tilde{d}_i}), \\ 
-2  \frac{3}{5} Y_{Q_i}^2 
 \left( {\alpha_1 \over 4 \pi} \right) 
 \left( {F_{\phi}\over 2 M_{H_C}} \right)
 M_1^* &\times&  f(M_1, m_{\tilde{u}_i}, m_{\tilde{d}_i}), 
\end{eqnarray}
respectively. 
Here, $Y_{Q_i}=1/6$ is a hyper charge for the superfield $Q_i$. 

Now, putting all corrections together, 
 the dimension five operator is described as 
\begin{eqnarray}
{\cal L}_{d=5} = \frac{x}{2 M_{H_C}} Y_{ikl} 
 (\psi_{Q_i} \psi_{Q_i}) (\tilde{Q}_k \tilde{L}_l), 
\end{eqnarray} 
where the factor $x$ is given by 
\begin{eqnarray}
x = 1+ 
 \frac{4}{3} \left( \frac{\alpha_3}{4 \pi} \right) 
 F_\phi M_3^* f_3 + 
 \frac{3}{2} \left( \frac{\alpha_2}{4 \pi} \right) 
 F_\phi M_2^* f_2 
 - \frac{1}{30} \left( \frac{\alpha_1}{4 \pi} \right)  
 F_\phi M_1^* f_1 . 
\end{eqnarray}
Here, $f_j = f(M_j, m_{\tilde{u}_i}, m_{\tilde{d}_i})$. 
In usual soft SUSY breaking mass spectrum, 
 soft masses for gauginos and sfermions are 
 of the same order, 
 and the loop function is characterized 
 by the scale of soft SUSY breaking masses, 
 $f_j \sim 1/M_{SUSY}$.
Thus, we find 
\begin{eqnarray} 
 x \sim  1  +  \frac{1}{M_{SUSY}}
  \sum_{i}  e^{-i \varphi_i} {\alpha_i \over 4 \pi} F_\phi ,
 \label{coeff} 
\end{eqnarray}
where $\varphi_i$ is a phase of each gaugino mass,
 $M_i =|M_i| e^{i \varphi_i}$.
Non-zero relative phases among gaugino masses 
 cause CP violation, and they are severely 
 constrained by current experiments on 
 electric dipole moments \cite{EDM}. 
Thus, in the following, we assume the universal phase, 
 $\varphi=\varphi_1=\varphi_2=\varphi_3$, 
 for all gaugino masses. 

As mentioned above, we consider the AMSB scenario, 
 where gauginos and sfermions obtain soft SUSY breaking masses 
 through superconformal anomaly such as \cite{AMSB1} \cite{AMSB2} 
\begin{eqnarray}
\label{gauginomass}
M_i &=& {\alpha_i \over 4 \pi} b_i F_\phi, \\ 
\tilde{m}_i^2 &=& 2 c_i 
 \left({\alpha_i \over 4 \pi}\right)^2 b_i |F_\phi|^2, 
\label{scalarmass}
\end{eqnarray}
respectively, 
 where $\alpha_i$ is the gauge coupling, 
 $b_i$ is a beta function coefficient, 
 and $c_i$ is a quadratic Casimir for each gauge interaction. 
Theses formulas imply that 
 $\frac{\alpha_i}{4 \pi} F_\phi \sim M_{SUSY}$ in Eq.~(\ref{coeff})
 and, therefore, 
 the dimension five operators induced by 1-loop
 corrections can be comparable to those at the tree level. 
This is a new contribution for proton decay processes we found. 
Note that the phase $e^{i \varphi}$ has an important role, 
  so that proton decay rate can be enhanced or suppressed 
  according to its value. 

If we assume the pure AMSB scenario, 
 soft SUSY breaking mass spectrum is fixed 
 and we obtain definite results  
 for the 1-loop induced dimension five operators. 
Especially, in this case, the phase $e^{-i \varphi}$ 
 is canceled by the phase of $F_\phi$ 
 and the sign of the new contribution is fixed 
 by the sign of the beta function coefficients. 
However, note that, unfortunately, the pure AMSB scenario 
 is obviously excluded, because it predicts 
 slepton squared masses being negative, 
 the so-called ``tachyonic slepton'' problem. 
There have been many attempts to solve this problem 
 by extending the pure AMSB model 
 and taking into account additional positive contributions 
 to slepton squared masses at tree level \cite{AMSB1} \cite{tree1} \cite{tree2} 
 or at quantum level \cite{PR} \cite{quantum}. 
Usually, these extended models still keep $F_\phi$  
 as an order parameter of SUSY breaking 
 and soft masses are characterized by $M_{SUSY} \sim 0.01 F_\phi$ 
 with 1-loop suppression factor $0.01$. 
We call these a class of models with AMSB. 
For these models, the phase of gaugino mass is, in general, 
 not necessary to be the same as the one of $F_\phi$ 
 and their relative phase can remain as a physical phase 
 in Eq.~(\ref{coeff}). 
Therefore, there is a possibility that the dimension five operators 
 induced by 1-loop corrections cancel those at the tree level 
 if we can choose an appropriate relative phase and 
 a soft mass spectrum. 

In conclusion, we have considered the effective superpotential 
 which leads to dimension five operators relevant 
 to the proton decay. 
In supergravity, the effective superpotential also leads 
 to soft SUSY breaking terms among sfermions, 
 dimension four operators. 
These terms originate from the $B$-term of color triplet Higgs bosons. 
The dimension four operators 
 can induce the dimension five operators 
 through 1-loop diagrams dressed by gauginos. 
In a class of models with AMSB, 
 the 1-loop contributions 
 are found to be comparable to the dimension five operators 
 at the tree level, 
 because the gravitino mass is two orders of magnitude 
 larger than the typical scale of soft SUSY breaking masses. 
Therefore, the proton decay rate can be enhanced or suppressed, 
 depending on the universal phase 
 of the gaugino masses and soft mass spectrum. 

We have discussed on the dimension five operators 
 induced by 1-loop corrections, 
 which can be comparable to those at the tree level. 
Our discussion is applicable to other effective superpotentials 
 appearing in some particle physics models. 
An example is an effective superpotential 
 which provides a light Majorana mass for neutrino 
 through the see-saw mechanism \cite{see-saw}, 
\begin{eqnarray} 
 W_{eff} =  \frac{(LH)(LH)}{M_R} , 
\end{eqnarray} 
where $M_R$ is the scale of right-handed Majorana neutrino mass. 
In the same manner as the above, 
 the superpotential leads to soft SUSY breaking term, 
\begin{eqnarray} 
 {\cal L}_{soft} = - \frac{F_\phi}{M_R} (\tilde{L} H)(\tilde{L}H). 
\end{eqnarray} 
Again, 1-loop corrections dressed by wino and bino 
 can be comparable to the tree level one 
 and resultant Majorana mass is altered from the one 
 at the tree level. 
If a phase of gaugino masses is different from that of $F_\phi$, 
 the phase additionally contributes to the Majorana phase 
 in the neutrino mixing matrix.

\vskip 1cm

\leftline{\bf Acknowledgments}
N.H. is supported by Alexander von Humboldt Foundation, 
and would like to thank M. Lindner for fruitful discussions. 
This work is supported in part by Scientific Grants from 
 the Ministry of Education and Science in Japan
 (Grant Nos.\ 15740164, 16028214, and 16540258). 

\vspace{.5cm}

%
%

%
\end{document}